\documentclass[12pt]{article}
\usepackage[dvips]{graphicx}

\newcommand{\be}{\begin{equation}}
\newcommand{\ee}{\end{equation}}
\newcommand{\cR}{\mathcal{R}}
\newcommand{\mn}{{\mu\nu}}
\newcommand{\tr}{\ensuremath{\mathop\mathrm{tr}}}
\newcommand{\duF}{\tilde{F}}
\newcommand{\ve}{\varepsilon}
\newcommand{\cn}{\ensuremath{\mathop{\mathrm{cn}}}}

\newcommand{\ec}{\ensuremath{\varepsilon_{c}}}
\newcommand{\Bk}{\ensuremath{\Sigma}}

\newcommand{\fR}{\ensuremath{\mathcal{P}}}

\begin{document}
\title{
  \begin{flushright} \small
   DTP--MSU/01-14 \\
   hep-th/0111099
  \end{flushright}
\vspace{1.cm}
\textbf{Non-Abelian Born--Infeld Cosmology}\footnote{Supported
by RFBR} }


\author{V.V. Dyadichev\footnote{Email: vlad@grg1.phys.msu.su}\,,
        D.V. Gal'tsov\footnote{Email: galtsov@grg.phys.msu.su}\,,
        A.G. Zorin\footnote{Email:  zrn@aport.ru}\,,\\[2mm]
    \sl Department of Theoretical Physics,\\
    \sl Moscow State University, 119899, Moscow, Russia\\[3mm]
        and M.Yu.\ Zotov\footnote{Email: zotov@eas.sinp.msu.ru}\\[2mm]
    \sl Skobeltsyn Institute of Nuclear Physics,\\
    \sl Moscow State University, 119899, Moscow, Russia
}

\date{\today}
\maketitle

\begin{abstract}
We investigate  homogeneous and isotropic cosmological solutions
supported by the SU(2) gauge field governed by the Born--Infeld
lagrangian. In the framework of the
Friedmann--Robertson--Walker cosmology with or without the
cosmological constant~$\lambda$, we derive dynamical systems
that give a rather complete description of  the space of
solutions. For $\lambda=0$ the effective equation of state
$\ve(p)$ is shown to interpolate between $p=-\ve/3$ in the
regime of the strong field and $p=\ve/3$ for the weak field.
Correspondingly, the Universe starts with zero acceleration and
gradually enters the decelerating regime, asymptotically
approaching the Tolman solution.
\end{abstract}

\vskip 0.3cm \indent \hskip 0.5cm
PACS numbers: 04.20.Jb, 04.50.+h, 46.70.Hg

\newpage
\section{Introduction}
It is tacitly assumed that the large scale massless Yang--Mills
(YM) fields, which could exist in the early Universe before
phase transitions, do not play any significant role in
cosmology. Partially this is related to the scale
invariance of the  YM lagrangian implying that primary  YM
excitations would be diluted during inflation. String theory
suggests the Born--Infeld (BI) type modification of the YM
action which breaks the scale invariance thus canceling the
above objection. Therefore, it seems reasonable to investigate the YM
cosmology with the non-Abelian Born-Infeld action.
In the existing literature one finds several
papers discussing cosmological models with the U(1) BI matter
\cite{GaBr00, Al90, Vo01}. Such models are necessarily anisotropic
(or inhomogeneous), since there is no homogeneous and isotropic
configurations of the classical U(1) field.
\emph{Non-Abelian} BI (NBI) cosmology was not
studied so far. Here we investigate this
problem in the framework of the traditional Friedmann-Robertson-Walker (FRW)
cosmology.

A notable property of the SU(2) Yang--Mills field is that it
\emph{admits} homogeneous and isotropic ($G_6$-invariant)
configurations. Indeed, the energy can be distributed between
three color components in such a way that the resulting
stress-tensor is compatible with the maximal symmetry of the
three-dimensional space. In the case of~$S^3$, the
corresponding ansatz was given in \cite{CeJa78,He82,VeDa89}, two other
cases (hyperbolic and flat) were treated in \cite{GaVo91}. An
interesting feature of these configurations is that they
produce a varying with time Chern--Simons density,
which might lead to topological fermion non-conservation
\cite{GiSt94,Vo94}. As was shown in \cite{GaVo91}, in the case
of the~$S^3$ spatial geometry, the YM field generically oscillates
between two neighboring topological sectors of the YM theory,
while the limiting unstable static solution plays a role of the
cosmological sphaleron \cite{GiSt94,Di94}.
Here we do not discuss this issue
further but rather concentrate on the aspects related to
the scale invariance breaking in the NBI theory.
An immediate consequence is
that the equation of state  $p=\ve/3$ arising in the usual YM theory
\cite{GaVo91} (mimicking the photon gas) changes to
some more complicated equation which now allows for a negative
pressure. One could wonder whether an accelerating expansion
becomes possible, that is, whether the sum $\ve+3p$ can be negative. It
turns out that the lower limit of  acceleration which can be
achieved within the present model is precisely zero: when YM field
strength is much larger then the BI `critical field', the
equation of state is $p=-\ve/3$. Such a state equation
is typical for an averaged distribution of strings, where its
origin is quite simple: each string has a {\em one-dimensional}
equation of state $p=-\ve$, averaging over all directions in the
three-space gives $p=-\ve/3$. The deeper reason for this
coincidence is not fully clear for the moment, though apparently
it is related to the fact that the origin of the BI lagrangian lies in the
string theory.

As it was widely discussed recently \cite{Ha81,Ts97,Ts99,Pa99},
the definition of the NBI action is ambiguous. One can start
with the U(1) BI action presented either in the determinant form
\[
     S=\frac1{16\pi}\int \,\sqrt{-\det(g_\mn-F_\mn)} \, d^4x,
\]
or in the `resolved square root' form
\[
     S=\frac1{16\pi} \int\,
     \sqrt{1+\frac{F_\mn F^\mn}{2}-\frac{({\tilde F}_\mn F^\mn )^2}{16}}
     \;\sqrt{-g} \, d^4x.
\]
In the non-Abelian case, the trace over color indices must be
specified. One particular definition, suggested by Tseytlin~\cite{Ts97},
is a symmetrized trace. It prescribes a
symmetrisation over all permutations of the gauge matrices in an
expansion of the determinant in powers of the field strength
before the trace  is
computed. Inside the symmetrized series expansion, the gauge generators
effectively commute, so both the determinant and the square root
forms are equivalent (which property may serve as an alternative definition
of the trace operation). This property does not hold for other trace
prescriptions, e.g., an ordinary trace. In this latter case, it
is common to apply the trace to the square root form. Note,
that the string theory favors the symmetrized trace definition
once lower orders of the perturbation theory are taken into
account \cite{Ts97,HaTa97,Ts99}, while higher order corrections
seem to violate this prescription
\cite{ReSaTe01,Bi01,Be01,SeTrTr01}. Here we choose the `square
root/ordinary trace' lagrangian just for its simplicity. It is
worth noting that in other (somehow related) problem of
sphalerons in the NBI theory, discussed recently both in the
ordinary~\cite{GaKe99} and symmetrized trace~\cite{DyGa00}
versions, basic qualitative features of the solutions turned out to be the
same. We realize, however, that the distinction between
different trace prescriptions may become significant near the
cosmological singularity.

Fortunately, the dynamics of the scale factor in the SU(2)
Einstein--NBI theory (with ordinary trace)
decouples from the whole system of equations and can be fully
analyzed using the dynamical systems approach. We perform this analysis
for closed, spatially flat and open FRW models
both for zero and non-zero cosmological constants.
Given the solution for the scale factor,
the field equations for the YM field can be integrated in
terms of elliptic functions.


\section{$SU(2)\;\; G_6$-invariant connection}

We start with the action
\begin{equation}\label{BItrace}
S=-\frac{1}{4\pi}\int \left\{\frac1{4G}R + \beta^2(\cR -
1)\right\}\;\sqrt{-g}\, d^4x,
\end{equation}
where $R$ is the scalar curvature, $\beta$ is the BI critical
field strength, and the quantity
\[
\cR=\sqrt{1+\frac{1}{2\beta^2}F^a_\mn F_a^\mn
    - \frac{1}{16\beta^4}(\duF^a_\mn F_a^\mn)^2}
\]
corresponds to the `square root/ordinary trace' NBI lagrangian.
Throughout the paper we consider cosmological models
with the homogeneous and isotropic three-space, which will
be presented as usual:
\begin{equation}\label{G6metr}
dl^2 = dr^2 + \Bk^2(d\theta^2+\sin\theta^2d\phi^2),
\end{equation}
where  $\Bk$ is one of the functions $\sin r$, $r$ or $\sinh r$
depending on the type of the spacetime: closed, spatially flat
or open (labeled as $k=1,0,-1$ correspondingly).

In what follows, we deal with  the homogeneous and isotropic
configurations of the SU(2) YM field. Although the construction
of the SU(2) YM ansatz for all FRW models was
discussed earlier~\cite{GaVo91}, it is worth to reconsider
it, since in~\cite{GaVo91} the conformal invariance of the
YM lagrangian was explicitly used. In fact, the result obtained
in~\cite{GaVo91} was just
the $G_6$-invariant connection independent of the choice
of the action for the gauge field. Here
we rederive the $G_6$ ansatz without any appeal to
the scale invariance.

The homogeneous and isotropic  ansatz for the YM field can be obtained
starting with the Witten  spherically  symmetric ansatz \cite{Wi77} and
then extending the symmetry to the full $G_6$ group. The Witten
ansatz reads:
\begin{eqnarray}\label{Wittans}
A_\mu dx^\mu&=& \omega_0 T_r \, d t + \omega_1 T_r \, d r
      + \left[ K_2 T_\theta- (1- K_1) T_\phi \right] d\theta \nonumber\\
            & & + \left[(1-K_1) T_\theta +K_2 T_\phi \right]
                    \sin \theta \, d\phi,
\end{eqnarray}
where four functions $\omega_0$, $\omega_1$, $K_1$, and~$K_2$
depend on time~$t$ and the radial coordinate~$r$ (spherical
coordinates $r,\theta,\phi$ are understood), and the coordinate dependent
basis in the color space is introduced as follows
\begin{equation}\label{Su2gens}
T_r =  T_1 \sin \theta \cos \phi
     + T_2 \sin \theta \sin \phi
     + T_3 \cos \theta, \qquad
T_\theta = \partial _\theta T_r, \qquad
T_\phi \sin \theta = \partial_\phi T_r.
\end{equation}
These generators satisfy the standard  normalisation condition
and the commutation relations
\begin{equation}\label{Su2comm}
  \tr T_a T_b=\frac12 \delta_{ab},\qquad
  [T_a,T_b]=i\varepsilon^{abc}T_c.
\end{equation}

From  four functions entering the ansatz only three are
physical, while $\alpha(r,t)=\arctan(K_2/K_1)$ can be gauged
away by a gauge transformation preserving the $SO(3)$ symmetry.
In the following  it is convenient to choose a parametrisation
\begin{eqnarray}\label{Omegas}
K_1=f \cos \alpha, &\qquad & K_2=f \sin \alpha, \nonumber \\
\omega_0=\Omega_0+\dot\alpha, &\qquad & \omega_1=\Omega_1 +
\alpha'.
\end{eqnarray}
where a prime and a dot denote derivatives with respect to the
radial variable and time.

In order to extend the symmetry  to~$G_6$,
we consider the following gauge invariant tensor:
\begin{equation}\label{Pdef}
  P_\nu{}^\lambda=F^a_{\mu\nu} F^{\mu\lambda}_a.
\end{equation}
Since in the homogeneous and isotropic three-space the only
mixed second rank tensor is the Kronecker delta, one should have for
the spatial part of~$P_\nu{}^\lambda$:
\[
  P_i^j=P(t)\;\delta_i^j.
\]
Using (\ref{Wittans}--\ref{Omegas}, we obtain the system of
equations for $\Omega_{0,1}$, $f$,~and $\alpha$, which is
solved for the first three functions in terms of a single
function of time~$w(t)$, while~$\alpha$ remains arbitrary. The
following gauge leads to a maximal simplification of the field
strength:
\begin{eqnarray}\label{Funcs}
\Omega_0 =-\frac{\dot{w} \Bk  \sqrt{1-k \Bk ^2}}{1+\Bk ^2(w^2-k)}, \\
\Omega_1 = \frac{\Bk ^2 w (w^2-k)}{1+\Bk ^2(w^2-k)}, \\
f  = \sqrt{1+\Bk ^2(w^2-k)}, \\
\alpha  = \arctan \left(\frac{\Bk w}{\sqrt{1-k \Bk^2}}\right).
\end{eqnarray}
In this gauge, the field  tensor reads
\begin{eqnarray}\label{Fform}
\mathcal{F}&=&\dot{w}\left(T_r\,dt\wedge dr
     + T_\theta \Bk\,dt \wedge d\theta
     + T_\phi \Bk \sin \theta \,dt \wedge d\phi \right) \nonumber \\
     &&+ \Bk(w^2-k)\left(T_\phi\,dr \wedge d\theta
     - T_\theta \sin \theta\,dr \wedge d\phi
     + T_r \Bk  \sin \theta\,d\theta \wedge d\phi \right).
\end{eqnarray}

\section{FRW cosmology}

Consider the  FRW cosmological models parameterizing
 the spacetime metric as
\[
     ds^2 = N^2 dt^2 - a^2 \,dl^2.
\]
Substituting (\ref{Fform}) into Eq.~(\ref{BItrace}), we
obtain the following one-dimensional action:
\[
 S_1= \frac{1}{4\pi}
      \int dt \left[\frac{3}{2 G}\frac{a(k N^2-\dot{a}^2)}{N}
      - N a^3 \beta^2 (\cR - 1) \right],
\]
where now
\[
\cR=\sqrt{1 - \frac{3\dot{w}^2}{a^2 \beta^2 N^2}
            + \frac{3(w^2-k)^2}{a^4 \beta^2}
            - \frac{9\dot{w}^2(w^2-k)^2}{a^6 \beta^4 N^2}}.
\]

The lagrangian contains two dimensional parameters:  the Newton
constant~$G$ and the BI `critical field'~$\beta$.
After a coordinate rescaling $t \to  \beta^{-1/2} t $,
$a\to\beta^{-1/2}a$, which makes all quantities dimensionless,
the reduced action becomes
\[
S_1=\frac{1}{4\pi G \beta} \int dt
     \left\{\frac32\frac{a(kN^2-\dot{a}^2)}{N}
     -g Na^3 \left[\sqrt{(1-K^2)(1+V^2)}- 1\right]\right\},
\]
where
\begin{equation}\label{VKdef}
     K=\frac{\sqrt 3 \dot{w}}{aN}\quad \mbox{and }\quad
     V=-\frac{\sqrt 3 (w^2-k)}{a^2},
\end{equation}
and $g=\beta G$ is the remaining dimensionless coupling constant.

Variation of the action over $N$ gives a constraint equation;
after obtaining it, we fix the gauge $N=1$. In this gauge the
constraint equation reads
\begin{equation}\label{Nconstr}
     g a^2 \left(\fR - 1 \right) - \frac32 (\dot{a}^2 + k) = 0,
\end{equation}
where
\begin{equation}\label{Rdef}
  \fR = \sqrt{\frac{1+V^2}{1-K^2}}.
\end{equation}
This expression can be rewritten in the standard form
\begin{equation}\label{habl}
\frac{\dot a^2}{a^2}+\frac{k}{a^2}=\frac{8\pi G }{3}\ve,
\end{equation}
where the energy density is given by
\begin{equation}\label{epsdef}
\ve=   \ec \left(\fR -1\right),
\end{equation}
with $\ec=\beta/4\pi$ playing a role of the BI critical energy density.

Variation of the action with respect to~$a$ gives the acceleration equation
\begin{equation}\label{aeq}
          \ddot{a}
     =\frac{ga}{3}\fR
     -\frac{\dot{a}^2+k+2ga^2}{2a}+\frac{2ga}{3}\fR^{-1}.
\end{equation}
Again, rewriting it in the standard form (using~(\ref{habl}))
\begin{equation}\label{acc}
     \frac{\ddot{a}}{a} = -\frac{4\pi G}{3}(\ve+3p),
\end{equation}
we can read off the pressure
\begin{equation}\label{pressdef}
     p=\frac13 \ec\left(3-\fR
       -2\fR^{-1}\right).
\end{equation}
Now comparing (\ref{epsdef}) and (\ref{pressdef}) we obtain the
following equation of state
\begin{equation}\label{state}
     p=\frac{\ve}{3}\frac{(\ec -\ve)}{(\ec+\ve)}.
\end{equation}
It is worth noting that the BI critical energy density corresponds to
the vanishing  pressure. For larger energies the pressure
becomes negative, its limiting value is $p=-\ve/3$. In the
opposite limit $\ve\ll \ec$ one recovers the hot matter equation
of state $p=\ve/3$, reflecting the scale invariance of the YM
action to which the NBI action reduces at low energies.

Finally, the variation over $w$ gives the YM (NBI) equation
\begin{equation}\label{weq}
     \ddot{w}= 2 \frac{w (k-w^2)}{a^2}
     \left(\frac{1-K^2}{1+V^2}\right) + \frac{2 \dot{a} \dot{w}}{a}
     \left(\frac12-\frac{1-K^2}{1+V^2}\right).
\end{equation}
 From this one can derive the following evolution equation
 for  the energy density:
\begin{equation}\label{drho}
     \dot {\varepsilon } =-2\frac{ \dot{a}}{a}\,\frac{\varepsilon
     \left (\varepsilon+2 \ec\right )}{\varepsilon+ \ec },
\end{equation}
which can be easily integrated to give
\begin{equation}\label{aepsC}
     a^4(\varepsilon+2\ec)\varepsilon={\rm const}.
\end{equation}
 From this relation one can see that the behavior of the NBI
field interpolates between two patterns: 1)~for large energy
densities ($\varepsilon\gg \ec $) the energy density scales as
$\ve\sim a^{-2}$;
2)~for small densities $\varepsilon\ll \ec$  one has a radiation
law $\varepsilon \sim \ a^{-4}$.


\section{Spacetime evolution}

Using the constraint equation (\ref{Nconstr}), one can
express~$\ve$ in terms of the scale factor~$a$ and its
derivative. Substituting it into Eq.~(\ref{aeq}) we
obtain a decoupled equation governing the spacetime evolution:
\begin{equation} \label{evol.eqn}
     \ddot a = -\frac{2 g a (\dot a ^2+k)}{2 g a^2 + 3 (\dot a^2 + k)}.
\end{equation}
The condition of positivity of the energy density in (\ref{Nconstr})
defines a boundary in the phase space:
\begin{equation}\label{domain}
     \dot a^2 > -k.
\end{equation}

Let us analyze the evolution equation~(\ref{evol.eqn}) by means
of the dynamical systems tools.  For this, it is convenient to
introduce $b = \dot a$ and pass to another independent variable~$\tau$, such
that $dt=h\,d\tau$, where $h = 2 g a^2 + 3 (b^2 + k)$. This leads
to the following dynamical system:
\begin{equation}\label{dynsys}
     a' = h b,  \qquad  b' = -2 g a (b^2  + k),
\end{equation}
where  primes denote derivatives with respect to~$\tau$. Notice
that the system~(\ref{dynsys}) is invariant under a reflection
$(a\to-a, \, b\to-b)$, so further, without loss of generality, we
discuss the expanding solutions.

First of all we
establish that the dynamical system~(\ref{dynsys}) admits a
first integral. Considering~$a$ as a function of~$b$, one can
write:
\[
 \frac{da}{db} = -\frac {h b}{2 a g (b^2 + k)}.
\]
This equation can be easily integrated to give
\begin{equation}\label{abcurve}
     3 \left(b^2 + k \right)^2 + 4 g a^2 \left(b^2 + k \right) = C,
\end{equation}
with some constant $C$. From this relation one can see that all
solutions inside the physical boundary~(\ref{domain}) cross the
line $a=0$, so that the singularity is unavoidable. Another
observation is that, unlike the standard hot FRW cosmology,
$\dot{a}$ remains finite while~$a$ tends to zero.

\begin{figure}
\centering
\includegraphics[width=11cm]{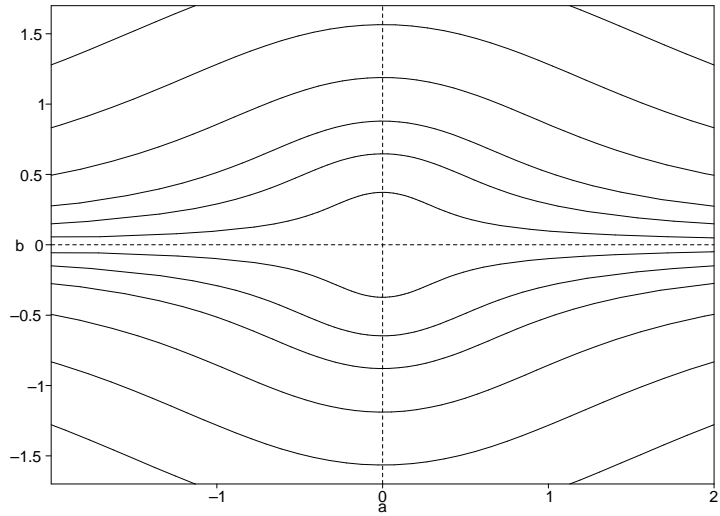}
\caption{Phase portrait of the dynamical system for the
spatially flat NBI universe.} \label{fig:dynflat}
\end{figure}

\begin{figure}
\centering
\includegraphics[width=11cm]{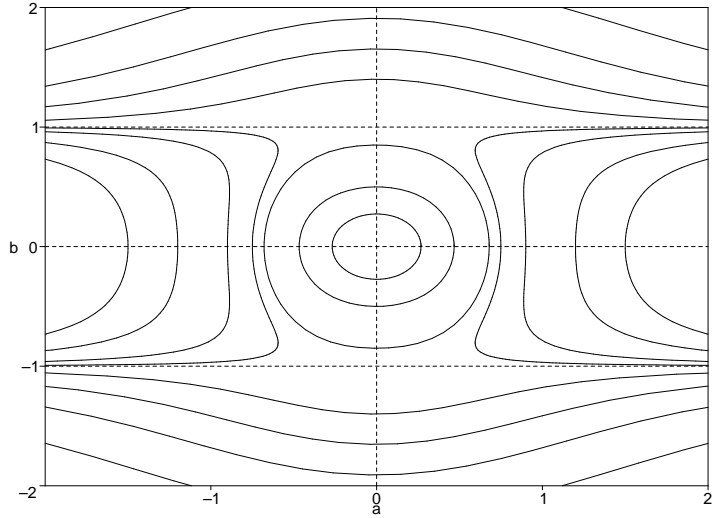}
\caption{Phase portrait of the dynamical system for the open NBI
universe. Physical domain in this case is $-1<b<1$.}
\label{fig:dynopen}
\end{figure}

Though the integral~(\ref{abcurve}) allows one to draw the
complete phase portraits of~(\ref{dynsys}), let us take a look
at the stationary points of the  dynamical system. These are
different for different~$k$:
\begin{itemize}
\item $k=1$, closed universe.
The only singular point is $a=0$, $b=0$ which is a center with
the eigenvalues $\pm i\sqrt{6g}$, see the phase portrait in
Fig.~\ref{fig:dynclsd}.%
\footnote{In this and the following figures, solutions evolve
from left to right in the upper half-plane
as time changes from~$-\infty$ to~$\infty$, and from right to left
in the lower half-plane.}
All solutions are of an oscillating type:
they start at the singularity ($a=0$) and after a stage of
expansion  shrink to another singularity. It is easy to see
from~(\ref{abcurve}) that the functions~$a$ and~$\dot a$ remain
bounded for all solutions.

\item $k=0$, spatially flat universe. There is a singular line
$b=0$ each point of which represents a solution for an empty
space (Minkowski spacetime). This set is degenerate, and there
are no solutions that reach this curve for finite values of~$a$
(see Fig.~\ref{fig:dynflat} for the phase portrait). All solutions
in the upper half-plane after initial singularity expand infinitely.

A remarkable fact is that for this case one can write an exact
solution of~(\ref{dynsys}) (in an implicit form):
\[
     4 \sqrt{g} \, (t - t_0)
     = \sqrt{3} \left(\Omega - \arctan \Omega^{-1} + \pi/2 \right),
\]
where $ \Omega=\sqrt{2}\,a/\sqrt{\sqrt{a^4+C}-a^2} $. The metric
singularity is reached at $t=t_0$.

\item $k=-1$, open universe. There is a center at $a=0$, $b=0$
with the eigenvalues $\pm i\sqrt{6g}$, but it lies outside the
boundary of the physical region.  Other  singular points are
$(a=0,\;\; b=\pm 1)$ (Fig.~\ref{fig:dynopen}). These points are
degenerate and cannot be reached from any point lying in the
physically allowed domain of the phase plane. The only
solutions which start from them are the separatrices $b=\pm 1$
that represent (part of) the flat Minkowski spacetime in special
coordinates.

One can easily see that all solutions in the upper part of the physical
domain
$\dot a>1$ start from the singularity and then move to
 $a \to \infty$, $ \dot a \to 1$.
\end{itemize}

The global qualitative behavior of  solutions does not differ
substantially from that in the conformally
invariant YM field model, except near the singularity.
 One can find
the following  power series expansion   near the
singularity $a(0)=0$:
\begin{equation}\label{aexp}
     a(t) = b_0 t - \frac{b_0 g}{9} t^3
            +\frac{1}{270}\frac{b_0 g^2 (7 b_0^2 + k)}{(b_0^2 + k)} t^5
            + O(t^7), \quad t \to 0,
\end{equation}
where $b_0$ is a free parameter. Absence of the quadratic term
means that the Universe starts with zero acceleration. This is
what can be expected in view of the equation of state
$p\approx-\ve/3$ at high densities.

For large $a$, the dynamics of the system~(\ref{aeq}) approaches
that of the hot FRW models (for small energy densities one
recovers the equation of state of radiation).

\section{Cosmological constant}

Here we extend our analysis to the case of non-zero
cosmological constant $\lambda$. This leads to the following
one-dimensional action:
\[
     S_1 = \frac{1}{4\pi G \beta } \int dt
     \left[ \frac{3a}{2N} \left(k N^2 - \dot a^2 \right)
     + g N a^3 \left(1 - \sqrt{(1-K^2)(1+V^2)} -\frac{\lambda}{2g}\right)
     \right].
\]
Clearly, the equations governing   the gauge field dynamics~%
(\ref{weq}) are unaffected by the cosmological term. The metric
equations can be decoupled again using the constraint equation~(\ref{Nconstr}):
\begin{equation}\label{aeqlam}
 \ddot a = -\frac 13 a(2 g - \lambda)
           +\frac43\frac{g^2a^3}{[(2g- \lambda)a^2+ 3(\dot a^2 +k)]}.
\end{equation}
The physical region of the phase space is defined by positivity of the
energy of the gauge field (\ref{epsdef}), now the boundary being
\begin{equation}\label{ldomain}
     \dot a^2 - \frac \lambda 3  a^2 +k \ge 0.
\end{equation}
For the closed and spatially flat universes and a negative
cosmological constant the allowed domain coincides with the
whole plane $a, \dot a$. In other cases, there is a boundary
corresponding to the solution with zero gauge field, i.e.,
the (anti)de-Sitter space.

The cosmological constant gives rise to new types of solutions
including de-Sitter-like, which are non-singular.
However, in the case when there is a singularity, its structure
is determined by the leading term in the equation of state,
namely, by the BI pressure, and is thus unaffected by the
cosmological constant. The generic solution  near the
singularity satisfies the following series expansion
\begin{equation}\label{explam}
     a(t)= b_0 t - \frac{2g-\lambda}{18} b_0 \,t^3 +O(t^5),
     \quad t \to 0,
\end{equation}
where $b_0$ is a free parameter. Again, the Universe starts with zero
acceleration.

 To proceed further with the analysis of Eq.~(\ref{aeqlam}),
we rewrite it as the following two-dimensional dynamical system:
\begin{equation}\label{dynsyslam}
     a' = 3 \left[(2g - \lambda) a^2 + 3 (b^2 + k)\right] b,
     \qquad
     b' = \left[\lambda(4g - \lambda) a^2 - 3 (b^2 + k)(2g - \lambda)
          \right] a,
\end{equation}
where primes denote derivatives with respect to a new time variable
$\tau$ defined by $dt=3 [(2g - \lambda) a^2 + 3 (b^2 + k)]  d\tau $.
The system is symmetric  under a
simultaneous reflection of both variables, so  we
can think  about a half of the phase space.

Similarly to the previous section, one may treat~$b$ as a
function of~$a$ and thus obtain the first integral, which
completely defines the integral curves of~(\ref{dynsyslam}):
\begin{equation}\label{lphase}
     3 \left(b^2-\frac\lambda3 a^2+k \right)^2
     + 4 g a^2 \left(b^2-\frac\lambda 3 a^2+k \right)= C.
\end{equation}
Note that the terms in brackets are non-negative within the
allowed domain~(\ref{ldomain}).

\begin{figure}
\centering
\includegraphics[width=11cm]{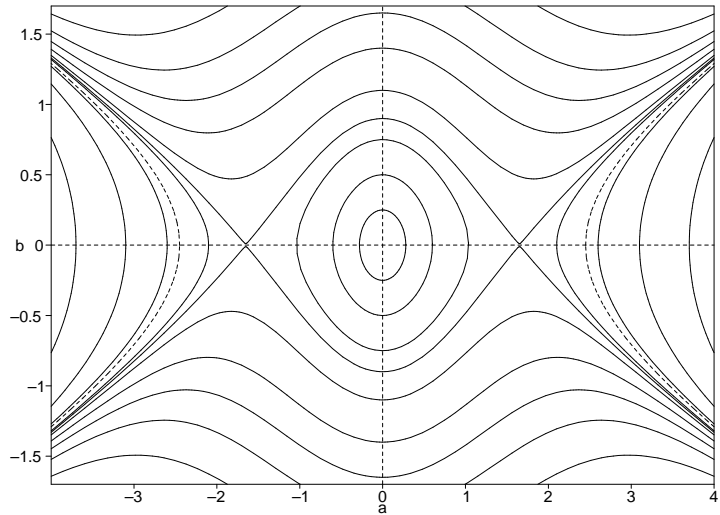}
\caption{Phase portrait of the dynamical
system~(\ref{dynsyslam}) for the closed NBI universe with a
cosmological constant $0<\lambda<2g$.  Physical domains are
bounded by hyperbolas plotted in dashed line.}
\label{fig:dynlam1}
\end{figure}

\begin{figure}
\centering
\includegraphics[width=11cm]{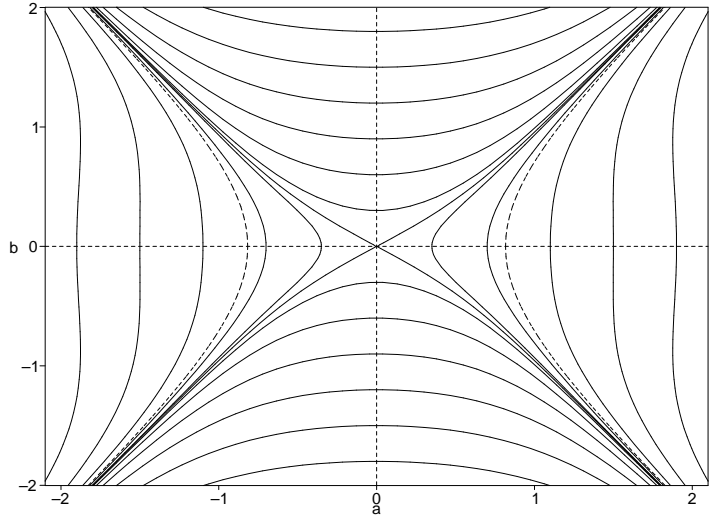}
\caption{Phase portrait of the dynamical system~(\ref{dynsyslam})
 for the closed NBI universe with a cosmological constant
$\lambda>2 g$.}
\label{fig:dynlam2}
\end{figure}

The structure of the solution space can be revealed using this
first integral. Let us start describing singular points of the
dynamical system for different~$k$.
\begin{itemize}
\item $k=1$, closed universe.  For a negative cosmological
constant the only singular point is $(a=0,b=0)$, which is a
center with eigenvalues $\xi = \pm 3 i \sqrt{3(2g-\lambda)}$.
All phase space trajectories are deformed circles, and all
solutions start from the singularity and reach the
singularity in the future.

For a positive cosmological constant, the allowed domain is bounded by
a hyperbola $b^2-\frac \lambda  3 a^2>-1$ which
represents the de-Sitter space.

For relatively  small~$\lambda$ the point $(a=0, b=0)$
remains a center. At the same time, another pair of singular
points appears:
\begin{equation}\label{clsp}
a =\pm\sqrt{\frac{3(2 g-\lambda)}{\lambda (4 g- \lambda )}},
\qquad b=0.
\end{equation}
For $\lambda<2g$ they lie within  the allowed domain and
correspond to the static Einstein universe. These are saddle
points with eigenvalues
\[
     \xi = \pm \frac{6 \sqrt{6}\,g}{\lambda}
               \sqrt{\frac{\lambda(2g-\lambda)}{4g-\lambda}}\,.
\]
Entering them separatrices correspond to either  a
solution developing from the singularity into the static universe,
or one rolling down from an infinite radius  to the static
universe (Fig.~\ref{fig:dynlam1}). These separatrices divide the
phase space into the domains containing different types of generic
solutions. Near the origin, all solutions evolve from an initial
to a final singularity. In the region
near the physical boundary, the solutions are
non-singular and evolve for an infinite time first shrinking to
some finite value of the scale factor and then ever expanding.
Generic solutions of the third type possess  a singularity but
are non-periodic: the universe expands forever.

With $\lambda$ further increasing, the saddle point approaches
the origin and finally, when $\lambda>2 g$, swallows it. The
character of the singular point $(a=0,b=0)$ changes---it
becomes a saddle point ((Fig.~\ref{fig:dynlam2}).
The separatrices entering it are solutions of an
inflationary type. They divide all generic solutions into
two classes: nonsingular de-Sitter-like (i.e.,
ever-expanding in the future and ever-contracting in the past)
and ever-expanding solutions with an initial singularity.
When $\lambda>4g$, the Eq.~(\ref{clsp})  has a real
singular point  lying outside the physical region.

\item $k=0$, spatially flat universe.
The only singular point  is the origin $(a=0,b=0)$ which now is
degenerate.
For $\lambda<0$ the allowed domain is the whole plane. All
solutions are of an oscillatory type  evolving from the
initial to the final singularity   for a finite time.

For a positive  cosmological constant the allowed
domain consists of the upper and lower parts of a cone
$b^2 - \frac \lambda 3 a^2>0$ whose boundary corresponds to the (part
of) de-Sitter space described in the inflating coordinates.
All solutions that lie within the allowed domain have an
initial singularity and are ever expanding in the future.

\item $k=-1$, open universe. The allowed domain~(\ref{ldomain})
lies outside of  either  an ellipse ($\lambda<0$), or  a
hyperbola  ($\lambda>0$).

In this case two types of singular points---the origin and
\begin{equation}
a =\pm\sqrt{\frac{3(\lambda-2 g)}{\lambda (4g-\lambda)}},
\qquad b=0
\end{equation}
(when they are real) lie outside the allowed domain and hence are
of no interest.

Another pair of singular points is $(a=0,b=\pm 1) $. These  are
degenerate (as in case  with zero cosmological constant).
Since they lie on the boundary of the allowed domain, the only
physical solutions  approaching them correspond  to zero energy
of the BI field  ((anti)-deSitter).

All  generic solutions within the allowed domain possess a
singularity and are either oscillating (for $\lambda<0$) or
ever expanding  (for $\lambda>0$).

\end{itemize}

\section{Dynamics of the gauge field }

One can easily see that the gauge field influences the metric
only through the quantity~\fR~(\ref{Rdef}),
related to the energy density.
This quantity obeys the   differential equation:
\begin{equation}\label{dR}
 \dot  \fR  =2 \frac{\dot a }{a}\left(\frac{1}{\fR} - \fR
 \right),
\end{equation}
which follows from (\ref{weq}) and does not
depend either on a detailed YM dynamics, or a particular gauge choice
(e.g., $N=1$, or $N=a$). This equation
can be integrated once, giving $\fR$ as a function of~$a$:
\begin{equation}\label{Radep}
  \fR=\sqrt{1+3\left(\frac{a_0}{a}\right)^4},
\end{equation}
where $a_0$ is an integration constant, and a numerical
coefficient was introduced for future convenience. This
equation gives us a possibility to fully describe the gauge
field dynamics. Recalling the definition of~$\fR$~(\ref{Rdef}),
one can separate the metric and the gauge field variables as follows:
\begin{equation}\label{w1eq}
 \frac{\dot{w}^2}{a_0^4-(k-w^2)^2} =\frac{ N^2a^2}{a^4+ 3
 a_0^4}.
\end{equation}
Since the right hand side of this equation is strictly positive,
we can find the domain of variation of the function~$w$. It
is symmetric with respect to $w\to -w$, except for the case
$a_0<1$, $k=1$, when~$w$ oscillates  near the value $w=1$ (or
$w=-1$)  without crossing the axis $w=0$.

The equation (\ref{w1eq}) can be solved in terms of the Jacobi
elliptic functions, similarly to the case of an ordinary YM
lagrangian~\cite{DoGa92}:
\begin{equation}\label{wsol}
  w(t)= \sqrt{k+a_0^2} \cn \left(\sqrt 2 a_0 (\tau -\tau_0),
  \frac{\sqrt{k+a_0^2}}{\sqrt{2} a_0}\right),
\end{equation}
where $\tau_0$ is the integration constant, and the argument to the
Jacobi function is defined as
\begin{equation}\label{cnarg}
  \tau(t)=\int^t \frac{a(t')N(t')\,dt'}{\sqrt{a^4(t')+3 a_0^4}}.
\end{equation}

\begin{figure}
\centering
\includegraphics[width=11cm]{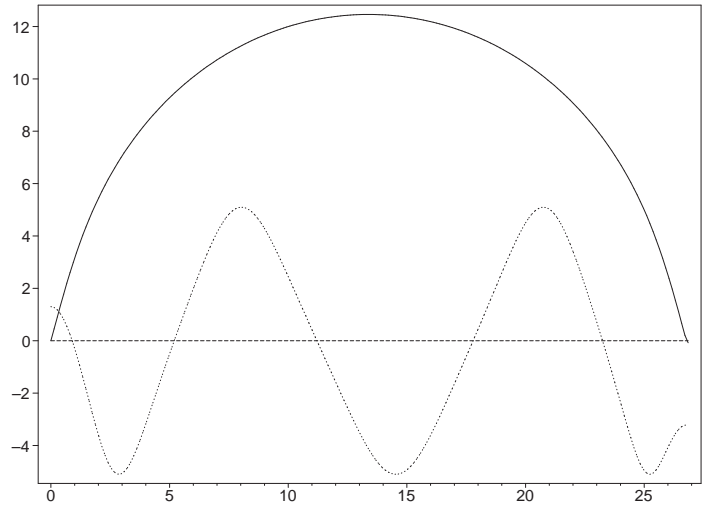}
\caption{Sample solution for the closed universe filled by the NBI YM field.
Dotted line---gauge field function~$w$, solid line---metric function~$a$.
The values of parameters are $g=\frac12$, $C=5$, $w_0=1.3$.}
\label{fig:clsol}
\end{figure}

\begin{figure}
\centering
\includegraphics[width=11cm]{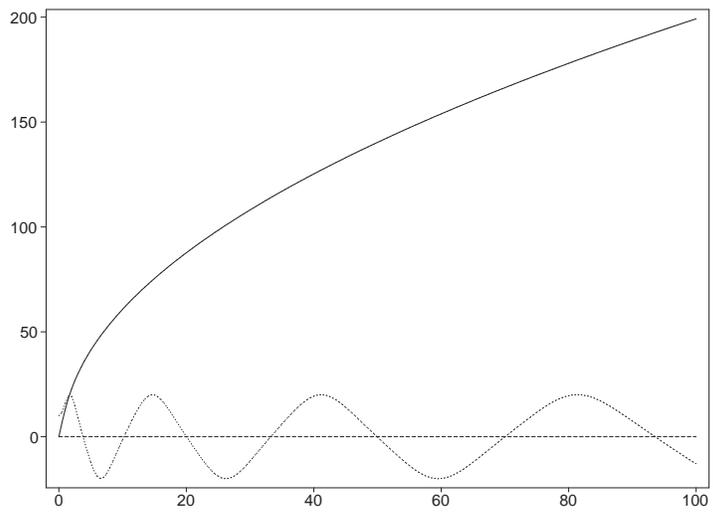}
\caption{Sample solution for the spatially flat universe with NBI YM
field. Dotted line---gauge field function~$w$, solid
line---metric function~$a$. }
\label{fig:flsol}
\end{figure}

A generic solution for $w(t)$ is of an oscillating type.
In the case $a_0<1$, $k = 1$, the solution,
oscillating near one of the  values $w=\pm 1$,
does not cross the line $w=0$. In
all other cases solutions oscillate around the origin. The
effective frequency of oscillations is determined by the rate
of growth of the argument~$\tau$. To compare the situation with the
ordinary EYM cosmology, we notice that in this case the solution
for~$w(t)$ is still given by~(\ref{wsol}), but with a different
definition of the phase variable~$\tau(t)$~\cite{DoGa92}
\[
     \tau_{EYM}(t)=\int^t \frac{N(t')}{a(t')} \, dt'.
\]
Clearly, the main difference with the ordinary EYM cosmology
relates to small values of~$a$. One can  see that near the
singularity ($a\to 0$) the YM oscillations in the NBI case slow
down, while in the ordinary YM cosmology the frequency remains
constant in the conformal gauge $N=a$, or tends to infinity in
proper time gauge $N=1$. The behavior of the YM field for closed
and spatially flat  models is illustrated
in Figs.(\ref{fig:clsol},\ref{fig:flsol}).

When the solution starts from the metric singularity, the gauge
field function has the following series expansion (the metric
is supposed to satisfy the expansion (\ref{explam})).
\begin{equation}\label{wexp}
  w=w_0+\frac{b_0\alpha}{6(k+b_0^2)} t^2+\left(\frac{g^2 b_0^2 w_0 (k-w_0^2)}
  {9(k+b_0^2)^2}+ \frac{b_0\alpha (\lambda- 2 g) }{216 (k+b_0^2)}\right)
  t^4+O(t^6), \quad t\to0,
\end{equation}
where $\alpha=\pm\sqrt{3(k+b_0^2)^2 - 4 g^2(k-w_0^2)^2}$,
$w_0$ is a free parameter (as well as the sign of~$\alpha$),
while~$b_0$ is a parameter from the expansion of the
metric~(\ref{explam}).

\section{Discussion}

In spite of considerable complications which the Born--Infeld
non-linearity introduces to the Einstein--Yang--Mills coupled
equations, the dynamics of the FRW NBI cosmology turned out to
be separable and admitting a rather complete analysis in terms
of the dynamical systems theory. This is due to the fact that the NBI
YM equation admits a first integral expressing the
variation of the energy density under
cosmological evolution. Thus one obtains a decoupled equation for
the scale factor both without or with the cosmological constant.
We have given a rather complete classification of possible
solutions for closed, spatially flat, and open universes in
these cases.

One of the intriguing features associated with the NBI theory
is the possibility of a negative pressure. Namely, the
pressure is negative when the field  energy density exceeds the
'critical' BI energy density. This  is still insufficient to
mimic an inflation, since the lowest value is only
$p=-\varepsilon/3$, which coincides with the equation of state
of the gas of Nambu-Goto strings in three spatial dimensions.
In the framework of the FRW cosmology, the equation of state of
the NBI YM field continuously interpolates from $p=-\varepsilon/3$
near the singularity to the `radiation' equation
$p=\varepsilon/3$ at large time. Correspondingly, the energy density is
evolving near the singularity according to the law~$a^{-2}$. The
FRW NBI universe starts with zero acceleration and achieves the
Tolman expansion regime, when the energy density is diluted
sufficiently to make the Born--Infeld non-linearity negligible.

Once the scale factor evolution is determined, one is able to
find an analytic solution for the YM function is terms of
elliptic functions. Qualitatively the YM dynamics remains the
same as in the usual YM theory, though we have observed that
near the singularity the oscillations of the YM field are
damped.

\section*{Acknowledgments}

The work was supported in part by the Russian Foundation for
Basic Research under grant 00-02-16306.


\newpage

\begin{figure}
\centering
\includegraphics[width=11cm]{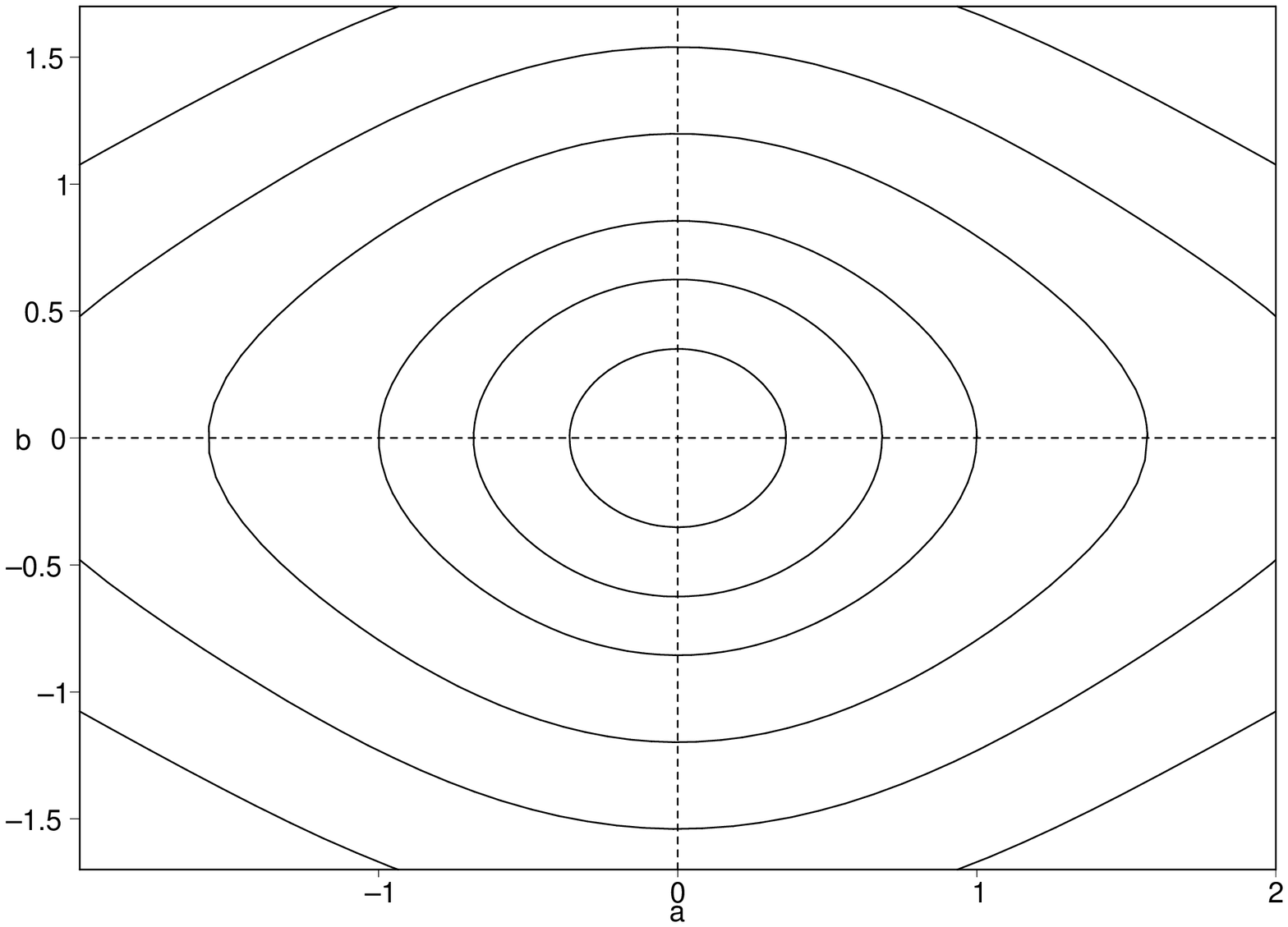}
\caption{Phase portrait of the dynamical system for the closed
NBI universe.} \label{fig:dynclsd}
\end{figure}

\end{document}